\documentclass[12pt,a4paper]{article}

\newcommand{\te}{\theta}
\newcommand{\la}{\lambda}
\newcommand{\pa}{\partial_y}

\begin{document}

\begin{flushright}
{}
\end{flushright}
\vspace{1.8cm}

\begin{center}
 \textbf{\Large Three-Spin Giant Magnons in $AdS_5 \times S^5$ }
\end{center}
\vspace{1.6cm}
\begin{center}
 Shijong Ryang
\end{center}

\begin{center}
\textit{Department of Physics \\ Kyoto Prefectural University of Medicine
\\ Taishogun, Kyoto 603-8334 Japan}  \par
\texttt{ryang@koto.kpu-m.ac.jp}
\end{center}
\vspace{2.8cm}
\begin{abstract}
From the Polyakov string action using a conformal gauge we
construct a three-spin giant magnon solution describing a long
open string in $AdS_5 \times S^5$ which rotates both in two
angular directions of $S^5$ and in one angular direction 
of $AdS_5$. Through the Virasoro constraints the string motion in
$AdS_5$ takes an effect from the string configuration in $S^5$.
The dispersion relation of the soliton solution is obtained as a
superposition of two bound states of magnons. We show that there is
a correspondence between a special giant magnon in $AdS_2$ and
the sinh-Gordon soliton.
\end{abstract} 
\vspace{3cm}
\begin{flushleft}
October, 2006
\end{flushleft}

\newpage
\section{Introduction}

According to the AdS/CFT correspondence \cite{MGW} the string theory in
$AdS_5 \times S^5$ should be dual to the $\mathcal{N} = 4$ super 
Yang-Mills (SYM) theory. The spectrum of certain string states matches
with the spectrum of dimensions of field theory operators in the SYM 
theory \cite{BMN,GKP}. There has been a mounting evidence that the 
spectrum of AdS/CFT is described by studying the multi-spin rotating
string solutions in $AdS_5 \times S^5$ \cite{FT,AT} and by analyzing
the Bethe equation for the diagonalization of the integrable spin
chain in the SYM theory \cite{MZ,BKS,NB,BS}. The direct relation between
both sides has been investigated at the level of effective action
\cite{MK,AAT} and the direct equivalence between the Bethe equation
for the spin chain and the classical Bethe equation for the classical
$AdS_5 \times S^5$ string sigma model has been shown from the
view point of integrability \cite{KMMZ,KZ}.

Recently, Hofman and Maldacena \cite{HM} have made a particular limit 
such that both the spin chain and the string effectively become
very long and constructed a rotating open string solution in $R \times
S^2$, namely, the giant magnon using the Nambu-Goto string action, which
is a particular case of the spiky string in $R \times S^2$ \cite{SR} 
that is the generalization of the spiky string in $AdS_3$ \cite{MKS},
and is identified with an elementary magnon 
excitation in the long spin chain. The dispersion relation 
between energy and angular momentum $J_1$ for the
giant magnon has been calculated to be equal to the strong 
't Hooft coupling limit of the dispersion relation for the spin chain
magnon that was derived by using the SU(2$|$2) $\times$ SU(2$|$2)
supersymmetry with a novel central extension \cite{NBT}. From the
equivalence between the string theory in $R \times S^2$ and the 
sine-Gordon theory \cite{KP,AM}, the giant magnon has been identified
with the sine-Gordon soliton and the scattering phase of two magnons has
been computed to be in agreement with the strong 't Hooft coupling limit
of the conjecture of \cite{AFS}. 

Analyzing the pole of the two-particle S-matrix \cite{ND} and exploiting 
the equivalence between the string theory in $R \times S^3$ and the 
complex sine-Gordon theory \cite{CDO}, the dispersion relation for the
two-charge dyonic giant magnon has been presented to be described in terms
of infinite $J_1$ and finite $J_2$ of angular momentum in an orthogonal
plane. This two-spin giant magnon is interpreted as a bound state composed
of $J_2$ magnons. From the Polyakov action the two-spin giant
magnon solution has been constructed for the string theory on  
$R \times S^3$ in the conformal gauge and the finite-size effects for
the string theory on  $R \times S^2$ in the uniform gauge have been 
studied \cite{AFZ}. The dispersion relation for the two-spin giant magnon
has been also derived from the Nambu-Goto action in the
static gauge and the one-loop superstring correction to the known folded
and circular two-spin string solutions has been analyzed in taking the
limit of infinite $J_1$ with finite $J_2$ \cite{MTT}.

For the string theory in the $\beta$-deformed
$AdS_5 \times S^5$ background
the two-spin giant magnon solution has been presented \cite{CGK,BR}.
Using the relation with the complex sine-Gordon theory, a family of closed
string solutions with two spins on $R \times S^3$ have been constructed
\cite{OS} such that they interpolate the rotating two-spin closed 
strings \cite{FT} and the dyonic giant magnons. 

Applying the dressing method \cite{ZM} to the SO(6) vector model 
describing strings in $R \times S^5$ the three-spin giant magnon solution
specified by infinite $J_1$ and finite $J_2$ and $J_3$ has been derived
\cite{SV} as a state consisting of two superposed, noninteracting,
two-charge bound states, one with a fixed momentum $\pi$
and the other with  an opposite momentum $-\pi$. There has been
a construction of the multi-spin giant magnon solution with
infinite $J_1$ and finite $J_2 = J_3$ \cite{BR}. 
From the Polyakov action for strings on $R \times S^5$ in the
conformal gauge the three-spin giant magnon solution in the
SU(3) sector has been constructed \cite{KRT} by generalizing the
Neumann-Rosochatius ansatz \cite{ART}, where it is interpreted
as a superposition of the bound state of $J_2$ magnons with a
total momentum $p_2$ and the bound state of $J_3$ magnons with a
different total momentum $p_3$. The S-matrix 
for bound states with an arbitrary
number of magnons in the SU(2) sector has been investigated
in both string and gauge theory sides \cite{CR}.

In the SL(2) sector using the Nambu-Goto action for strings in
$AdS_3 \times S^1$ \cite{MTT} and that with NS-NS B field \cite{WH},
the two-spin giant magnon solutions have been presented
in the static gauge. In order to take an attempt to extend the
SL(2) sector to the larger sector we will use
the Polyakov action for strings
in $AdS_3 \times S^3$ in the conformal gauge to construct a 
three-spin giant magnon with one angular momentum in $AdS_3$ and two
angular momenta in $S^3$. We will see that the two Virasoro constraints
connect the two subsectors SL(2) and SU(2) and make an important
role to determine the form of the rotating open string configuration.
We will discuss a relation between a giant magnon solution on
$AdS_2$ and the sinh-Gordon soliton.

\section{Three-spin giant magnons}

We consider a three-spin giant magnon in $AdS_5 \times S^5$ which has
one spin $S$ in $AdS_5$ and two spins $J_1$ and $J_2$ in $S^5$.
The relevant metric is that of $AdS_3 \times S^3$ part of 
$AdS_5 \times S^5$
\begin{equation}
ds^2 = - \cosh^2 \rho dt^2 + d\rho^2 + \sinh^2\rho d\varphi^2 + 
d\te^2 + \cos^2\te d\phi_1^2 + \sin^2\te d\phi_2^2, 
\end{equation}
so that the Polyakov string action in the conformal gauge becomes
\begin{eqnarray}
I &=& -\frac{\sqrt{\la}}{4\pi} \int d\tau d\sigma [ -\cosh^2 \rho(t'^2
- \dot{t}^2) + \rho'^2 - \dot{\rho}^2 + \sinh^2\rho( \varphi'^2 -
\dot{\varphi}^2 ) \nonumber \\
&+& \te'^2 - \dot{\te}^2 + \cos^2\te ({\phi'}_1^2 - \dot{\phi}_1^2 ) +
\sin^2\te ({\phi'}_2^2 - \dot{\phi}_2^2 ) ], 
\end{eqnarray}
where the dot and prime denote the derivatives with respect to $\tau$ and
$\sigma$ which ranges from $-\infty$ to $\infty$.
We make the ansatz for a rotating open string soliton
\begin{eqnarray}
t &=& \tau + h_1(y), \hspace{1cm} \rho = \rho(y), \hspace{1cm}
\varphi = \omega (\tau + h_2(y) ), \nonumber \\
\phi_1 &=& \tau + g_1(y), \hspace{1cm} \te = \te(y), \hspace{1cm}
\phi_2 = w(\tau + g_2(y)),
\label{ans}\end{eqnarray}
where $y = \sigma - v\tau$. The equations of motion for $\phi_1, \phi_2$ 
lead to
\begin{equation}
\pa g_1 = \frac{v}{1-v^2} \tan^2\te, \hspace{1cm} \pa g_2 = -
\frac{v}{1-v^2},
\end{equation}
which give the equation of motion for $\te$
\begin{equation}
(1-v^2)^2 \pa^2\te = \sin\te \cos\te \left( 1-w^2 - 
\frac{v^2}{\cos^4\te} \right).
\label{the}\end{equation}
The first integral of (\ref{the}) with an appropriate integration constant
is obtained by
\begin{eqnarray}
(1-v^2)^2 (\pa\te)^2 &=& \sin^2\te \left( 1-w^2 - \frac{v^2}{\cos^2\te}
\right) \nonumber \\
&=& (1-w^2) \tan^2\te (\alpha^2 - \sin^2\te)
\label{fit}\end{eqnarray}
with $\alpha=\sqrt{(1-v^2-w^2)/(1-w^2)}$, whose solution is given by
\begin{equation}
\sin\te = \frac{\alpha}{\cosh\beta y}, \hspace{1cm} \mathrm{for}
-\infty < y < \infty,
\label{tal}\end{equation}
where $\beta = \sqrt{1-v^2-w^2}/(1-v^2), 1-v^2-w^2\ge 0, 1-w^2\ge 0$.
The angle $\phi_2$ is also expressed as $\phi_2 =
w(\tau - v\sigma)/(1-v^2)$.
These expressions were presented in \cite{AFZ} by using the conformal
gauge supplemented by the static choice $t=\tau$
for the Polyakov action of strings in $R \times S^3$.

The equations of motion for $t$ and $\varphi$ are given by
\begin{eqnarray}
\pa[ (v + (1-v^2)\pa h_1) \cosh^2\rho ] = 0, \nonumber \\ 
\pa[ (v + (1-v^2)\pa h_2) \sinh^2\rho ] = 0,
\end{eqnarray}
which generate two conservation laws
\begin{equation}
\pa h_1 = \frac{1}{1-v^2}\left( -v + \frac{c_1}{\cosh^2\rho} \right),
\hspace{1cm} \pa h_2 = \frac{1}{1-v^2}\left( -v + 
\frac{c_2}{\sinh^2\rho}\right),
\label{twh}\end{equation}
where $c_1$ and $c_2$ are integration constants. We use the 
expressions in (\ref{twh}) to obtain the equation of motion for $\rho$
\begin{equation}
(1-v^2)^2 \pa^2\rho = -\sinh\rho \cosh\rho \left[ \omega^2
\left(1 - \frac{c_2^2}{\sinh^4\rho} \right) 
 - \left(1 - \frac{c_1^2}{\cosh^4\rho} \right) \right],
\label{rho}\end{equation}
which is compared with (\ref{the}).

We first express one constraint on the energy-momentum tensor of the
system $T_{\tau \sigma}$ in terms of $z=\sin \te$ as
\begin{eqnarray}
\frac{1}{v}(1-v\pa h_1) \pa h_1 \cosh^2\rho + (\pa \rho)^2 
-\frac{\omega^2}{v}(1-v\pa h_2) \pa h_2 \sinh^2\rho \nonumber \\
+ (\pa\te)^2 - \frac{1}{1 - v^2}\left( 1 - \frac{v^2}{1-v^2}
\frac{z^2}{1-z^2} \right)z^2 + \frac{w^2z^2}{(1-v^2)^2} = 0,
\end{eqnarray}
which combines with (\ref{fit}) to be
\begin{equation}
\frac{1}{v}(1-v\pa h_1) \pa h_1 \cosh^2\rho + (\pa \rho)^2 
-\frac{\omega^2}{v}(1-v\pa h_2) \pa h_2 \sinh^2\rho = 0.
\label{onv}\end{equation}
The other constraint $T_{\tau\tau} + T_{\sigma\sigma}=0$ now reads
\begin{eqnarray}
-[(1-v\pa h_1)^2 + (\pa h_1)^2 ] \frac{\cosh^2\rho}{1+v^2} + (\pa\rho)^2
+\omega^2[(1-v\pa h_2)^2 + (\pa h_2)^2 ] \frac{\sinh^2\rho}{1+v^2}
\nonumber \\
+ (\pa\te)^2 + (1-z^2)\left[ \frac{1}{1 + v^2}\left( 1 - 
\frac{2v^2}{1-v^2}\frac{z^2}{1-z^2} \right) + \frac{z^4v^2}
{(1-v^2)^2(1-z^2)^2} \right] + \frac{w^2z^2}{(1-v^2)^2} = 0,
\end{eqnarray}
which also turns out to be of the form
\begin{equation}
-[(1-v\pa h_1)^2 + (\pa h_1)^2 ] \frac{\cosh^2\rho}{1+v^2} + (\pa\rho)^2
+\omega^2[(1-v\pa h_2)^2 + (\pa h_2)^2 ] \frac{\sinh^2\rho}{1+v^2}
+\frac{1}{1+v^2} =0
\label{twv}\end{equation}
owing to (\ref{fit}). We require that the equation (\ref{onv}) should be
identically equal to (\ref{twv}). Eliminating the $(\pa\rho)^2$ term 
in (\ref{onv}) and (\ref{twv}) by subtraction we have one relation 
expressed in terms of $\pa h_1$ and $\pa h_2$. The substitution of 
(\ref{twh}) into this extracted relation yields 
\begin{equation}
c_1 - \omega^2c_2 = v.
\end{equation}

In what follows we will take a simple case $c_1 =v, c_2=0$.
From (\ref{ans}) and (\ref{twh}) with $c_1=v$ we have
\begin{equation}
\frac{dt}{d\tau} = \frac{\cosh^2\rho - v^2}{(1-v^2)\cosh^2\rho} > 0,
\end{equation}
which insures forward propagation in time. Thus the equations (\ref{onv})
and (\ref{twv}) become the same expression
\begin{eqnarray}
(1-v^2)^2(\pa\rho)^2 &=& \sinh^2\rho \left( 1 - \omega^2 - 
\frac{v^2}{\cosh^2\rho} \right) \nonumber \\
&=& (1-\omega^2) \tanh^2\rho \left( \sinh^2\rho + \frac{1 -v^2 -\omega^2}
{1 - \omega^2} \right),
\label{frh}\end{eqnarray}
which is indeed the first integral of the equation of motion for $\rho$
(\ref{rho}) and depends on the two
parameters $v$ and $\omega$ in the form similar to (\ref{fit}).
Since the $1-\omega^2 \le 0$ region is not allowed in view of the  
expression (\ref{frh}), there are two parameter regions
\begin{eqnarray}
\mathrm{A} &:& \hspace{1cm} 1-v^2 - \omega^2 \le 0, \; 1- \omega^2 \ge 0,
\nonumber \\
\mathrm{B} &:& \hspace{1cm} 1-v^2 - \omega^2 \ge 0, \; 1- \omega^2 \ge 0,
\end{eqnarray}
which are also expressed as A: $0 \le 1-\omega^2 \le v^2$ and B: 
$v^2 \le 1- \omega^2$.

For the region A the eq. (\ref{frh}) leads to 
\begin{equation}
\pa\rho = \pm \frac{\sqrt{1-\omega^2}}{1-v^2}\tanh\rho 
\sqrt{\sinh^2\rho - \tilde{\alpha}^2},\hspace{1cm} \tilde{\alpha} = 
\sqrt{\frac{v^2 + \omega^2 -1}{1 -\omega^2} },
\end{equation}
which can be integrated as
\begin{equation}
\frac{1}{\tilde{\alpha}} \sqrt{\sinh^2\rho - \tilde{\alpha}^2} =
\left\{ \begin{array}{rl} \tan\tilde{\beta}y, \hspace{1cm} & 
\mbox{for $0 \le y \le \frac{\pi}{2\tilde{\beta}}$, } \\
 -\tan\tilde{\beta}y, \hspace{1cm}
& \mbox{for -$\frac{\pi}{2\tilde{\beta}} \le y \le 0$ }  \end{array} 
\right.
\end{equation}
with $\tilde{\beta} = \sqrt{v^2 + \omega^2 -1}/(1 -v^2)$.
This solution lies within a finite range of $y$ and is 
expressed as 
\begin{equation}
\sinh\rho = \frac{\tilde{\alpha}}{\cos\tilde{\beta}y}, \hspace{1cm}
\mathrm{for} -\frac{\pi}{2\tilde{\beta}} \le y \le 
\frac{\pi}{2\tilde{\beta}}.
\end{equation}
At the boundaries of range $y=\pm \frac{\pi}{2\tilde{\beta}}$ 
the radial coordinate $\rho$ extends to the infinity, and
at $y=0$ it becomes the shortest value $\rho_0 = \sinh^{-1}
\tilde{\alpha}$, which is compared with the maximum value
$\te_{\mathrm{max}} = \sin^{-1}\alpha$ at $y=0$ for solution (\ref{tal}).
In the region A we cannot make a $\omega =0$ reduction that
corresponds to the string solution in $AdS_2 \times S^3$,
while the $\omega =0$ reduction is possible in the region B.

For the region B through an appropriate integration constant
the eq. (\ref{frh}) is similarly integrated as
\begin{equation}
\sinh\rho = 
\left\{ \begin{array}{rl} \frac{\hat{\alpha}}{\sinh\hat{\beta}y},
 \hspace{1cm} & 
\mbox{for $0 \le y < \infty$,} \\
-\frac{\hat{\alpha}}{\sinh\hat{\beta}y}, \hspace{1cm}
& \mbox{for $-\infty < y \le 0$, }  \end{array}\right.
\label{sol}\end{equation} 
where
\begin{equation}
\hat{\alpha}=\sqrt{ \frac{1 - v^2 -\omega^2}{1 -\omega^2}},
\hspace{1cm} \hat{\beta} = \frac{\sqrt{1 - v^2 -\omega^2}}{1-v^2}.
\end{equation}
At $y = 0 \; \rho$ extends infinitely to the boundary of $AdS_3$,
while at $y = \pm\infty \; \rho$ becomes zero to reach the origin of 
$AdS_3$. Since this solution is supported in the same infinite range
$-\infty < y < \infty$ as that for the solution (\ref{tal}), we will
analyze the string solution in the parameter region B, which describes an
open string on a plane.

The rotating open string is characterized by the following energy and
spins
\begin{eqnarray}
E &=& \frac{\sqrt{\la}}{2\pi}\int d\sigma \cosh^2\rho \left( 1+ 
\frac{v^2}{1-v^2}\tanh^2\rho \right), \nonumber \\
J_1 &=& \frac{\sqrt{\la}}{2\pi}\int d\sigma \cos^2\te \left( 1- 
\frac{v^2}{1-v^2}\tan^2\te \right), \nonumber \\
J_2 &=& \frac{\sqrt{\la}}{2\pi} \frac{w}{1-v^2}\int d\sigma \sin^2\te,
\nonumber \\
S &=& \frac{\sqrt{\la}}{2\pi}\frac{\omega}{1-v^2}\int d\sigma \sinh^2\rho,
\label{cha}\end{eqnarray}
where $J_1, J_2$ and $S$ are the spins associated with the $\phi_1,
\phi_2$ and $\varphi$ directions. They combine to yield a relation
\begin{equation}
E - J_1 = \frac{S}{\omega} + \frac{J_2}{w}.
\label{rel}\end{equation}
When the solutions (\ref{sol}) and (\ref{tal}) are plugged into the 
respective expressions in (\ref{cha}) we see that both $E$ and $J_1$ 
diverge owing to the effectively long open string configuration, while the
difference $E - J_1$ has no such IR divergence 
but contains the UV divergence.
The solution (\ref{tal}) is used to rewrite the $J_2/w$ term in 
(\ref{rel}) as a finite value \cite{AFZ,MTT}
\begin{equation}
\frac{J_2}{w} = \sqrt{J_2^2 + \frac{\la}{\pi^2}\alpha^2 },
\label{sqr}\end{equation}
where $\alpha$ is characterized by an angle difference between the two
endpoints of the open string as
\begin{equation}
\Delta \phi_1 = \int_{-\infty}^{\infty}dy \pa g_1(y) = 2\cos^{-1}
\sqrt{1- \alpha^2},
\end{equation}
which is identified with the magnon momentum and yields $\alpha = 
\sin\frac{\Delta\phi_1}{2}$. 

On the other hand the $S$ spin contribution
to the string energy is expressed through (\ref{frh}) and the change of
variable $z = \cosh\rho$ as 
\begin{equation}
\frac{S}{\omega} = \frac{\sqrt{\la}}{2\pi}\frac{2}{1 -v^2}\int_{\infty}^0
d\rho \frac{\sinh^2\rho}{\pa \rho} =
\frac{\sqrt{\la}}{\pi}\frac{1}{\sqrt{1- \omega^2}}\int_1^{\infty}dz
\frac{z}{\sqrt{z^2 - z_0^2}},
\label{som}\end{equation}
where $z_0 = v/\sqrt{1- \omega^2}$ and $\pa\rho = -\sqrt{1-\omega^2}
\tanh\rho\sqrt{\sinh^2\rho + \hat{\alpha}^2}/(1-v^2)$ for 
$0 \le y < \infty$.  However, this integration diverges 
because the rotating long string stretches to the boundary of $AdS_3$.
By introducing a cutoff $\Lambda$ to regulate the UV divergence 
we evaluate (\ref{som}) as
\begin{equation}
\frac{S}{\omega} = \frac{\sqrt{\la}}{\pi}\frac{1}{\sqrt{1- \omega^2}}
( \Lambda - \sqrt{1 - z_0^2} ).
\end{equation}
This divergence appeared when the giant magnon solution was derived by
analyzing the Nambu-Goto action in the static gauge for the string
with two spins $J_1$ and $S$ in $AdS_3 \times S^1$ \cite{MTT}.
Following the prescription of ref. \cite{MTT} we subtract the divergent
term to have a regulated value
\begin{equation}
\frac{S_{\mathrm{reg}} }{\omega} = -\frac{\sqrt{\la}}{\pi}
\sqrt{ \frac{1 - z_0^2}{1- \omega^2} },
\label{reg}\end{equation}
from which $\omega$ is obtained by 
\begin{equation}
\omega = \frac{|S_{\mathrm{reg}}|}{ \sqrt{S_{\mathrm{reg}}^2 +
\frac{\la}{\pi^2}( 1- z_0^2)} }.
\label{ome}\end{equation}
Combining (\ref{reg}) and (\ref{ome}) the $S$ spin contribution
is expressed as a magnon-like dispersion relation
\begin{equation}
\frac{S_{\mathrm{reg}} }{\omega} = - \sqrt{S_{\mathrm{reg}}^2 +
\frac{\la}{\pi^2}\hat{\alpha}^2 },
\end{equation}
which resembles (\ref{sqr}), where $\hat{\alpha}$ and $\alpha$ are 
similarly defined by using $\omega$ and $w$ respectively.
Here the parameter $\hat{\alpha}$ is 
characterized by a time difference between the two endpoints of the
open string
\begin{equation}
\Delta t = \int_{-\infty}^{\infty}
dy \pa h_1(y) = -2 \tan^{-1}\frac{\sqrt{1- \hat{\alpha}^2} }
{\hat{\alpha}},
\label{del}\end{equation}
which reduces to $\hat{\alpha}=\cos\frac{\Delta t}{2}$.  
Thus we have a dispersion relation for the three-spin giant magnon
in $AdS_3 \times S^3$
\begin{equation}
(E - J_1)_{\mathrm{reg}} = - \sqrt{S_{\mathrm{reg}}^2 +
\frac{\la}{\pi^2}\cos^2\frac{\Delta t}{2} } +
\sqrt{J_2^2 + \frac{\la}{\pi^2}\sin^2\frac{\Delta\phi_1}{2} },
\end{equation}
which is regarded as the energy of a superposition of a bound state
of $J_2$ magnons with momentum $\Delta\phi_1$ and a bound state of
$|S_{\mathrm{reg}}|$ magnons with momentum $\pi + \Delta t$.
Up to a negative sign this dispersion relation has the similar 
structure to that for the giant magnon with the three spins 
$J_1, J_2, J_3$ in $R \times S^5$ \cite{SV,KRT}.

\section{Giant magnons on $AdS_2$ and sinh-Gordon solitons}

Let us analyze the $AdS_3$ part of the three-spin giant magnon 
configuration. The $AdS_3$ space-time is parametrized by a complex 
two-component vector $Y_i = (Y_0, Y_1)$
\begin{equation}
Y_0 = \cosh\rho e^{it}, \hspace{1cm} Y_1 = \sinh\rho e^{i\varphi},
\label{coy}\end{equation}
which obeys $Y_i^{*}Y^i = -1, \; Y^i =\eta^{ij}Y_j$, with
$\eta^{ij}= \mathrm{diag}(-1,1)$. Alternatively, a real four-component
vector
\begin{equation}
n_i = (\cosh\rho \cos t, \cosh\rho \sin t, \sinh\rho \cos\varphi,
\sinh\rho \sin\varphi )
\end{equation}
parametrizes the $AdS_3$ space-time in such a way that $n_in^i = -1$
and $n^i =\eta^{ij}n_j$ with the flat $R^{2,2}$ metric 
$\eta^{ij} = \mathrm{diag}(-1,-1,1,1)$.
The reduced Virasoro constraint (\ref{twv}) for the string motion in
$AdS_3$ can be expressed in a compact form as
\begin{equation}
\dot{n}_i\dot{n}^i + n'_in'^i = -1.
\label{sun}\end{equation}

Here in order to capture a fascinating feature of the giant magnon on 
$AdS_3$ we try to compute a combination
\begin{equation}
\dot{n}_i\dot{n}^i - n'_in'^i = -\cosh^2\rho( \dot{t}^2 - t'^2 )
+ \dot{\rho}^2 - \rho'^2 + \sinh^2\rho( \dot{\varphi}^2 - \varphi'^2 ).
\label{dif}\end{equation}
By substituting the solution (\ref{sol}) into the first equation in
(\ref{twh}) and integrating it we express the time coordinate $t$ as
\begin{equation}
t- \tau = h_1 = -\tan^{-1}\left(\frac{\sqrt{1-\hat{\alpha}^2}}
{\hat{\alpha}}\tanh\hat{\beta}y \right) + k,
\label{tta}\end{equation}
which reproduces (\ref{del}). If the integration constant $k$ is
chosen to be zero, it is convenient to express (\ref{tta}) as
\begin{equation}
\cot(t - \tau) = -\frac{\hat{\alpha}}{\sqrt{1- \hat{\alpha}^2}}
\coth\hat{\beta}y.
\label{tat}\end{equation}
The eq. (\ref{tat}) combines with (\ref{sol}) to yield
\begin{equation}
\cosh\rho = 
\left\{ \begin{array}{rl} \frac{\sqrt{1- \hat{\alpha}^2}}{\sin(t-\tau)},
\hspace{1cm} & \mbox{for $0 \le t - \tau \le \hat{h}_1 < 
\frac{\pi}{2}$} \; (-\infty < y \le 0),  \\
-\frac{\sqrt{1- \hat{\alpha}^2}}{\sin(t-\tau)}, \hspace{1cm}
& \mbox{for -$\frac{\pi}{2} < -\hat{h}_1 \le t - \tau  \le 0$} \; 
( 0 \le y < \infty ),  \end{array}\right.
\label{cor}\end{equation} 
where $\hat{h}_1 = \tan^{-1}(\sqrt{1-\hat{\alpha}^2}/\hat{\alpha})$.
By substituting the derivatives of (\ref{sol}) and (\ref{tat}) with 
respect to $\tau$ and $\sigma$  into (\ref{dif}) 
and taking account of (\ref{cor})
and $\pa h_2=-v/(1-v^2)$ we find a concise expression
\begin{equation}
\dot{n}_i\dot{n}^i - n'_in'^i = - \left( 1 + \frac{2(1-\omega^2)
\hat{\alpha}^2}{1-v^2} \frac{1}{\sinh^2\hat{\beta}y} \right).
\label{non}\end{equation}
It can be checked that the constraint (\ref{sun}) is indeed satisfied by
the explicit relations (\ref{tat}) and (\ref{cor}).
Alternatively, instead of the explicit relations we directly use
(\ref{twh}) to express (\ref{dif}) as
\begin{eqnarray}
\dot{n}_i\dot{n}^i - n'_in'^i &=& -\cosh^2\rho - \frac{2v^2}{1- v^2}
\sinh^2\rho \nonumber \\
&+& (1- v^2)\left[ \frac{v^2}{(1-v^2)^2}\frac{\sinh^4\rho}{\cosh^2\rho}
- (\pa\rho)^2 + \frac{\omega^2}{(1-v^2)^2}\sinh^2\rho \right],
\end{eqnarray}
which reduces to (\ref{non}) through the substitution of (\ref{sol}).

If we define a scalar field $\phi$ as
\begin{equation}
\cosh2\phi = - ( \dot{n}_i\dot{n}^i - n'_in'^i ),
\end{equation}
whose $\phi$ is real bcause of comparing $\cosh2\phi >1$ with 
(\ref{non}), we obtain 
\begin{equation}
\sinh\phi = \sqrt{ \frac{1- v^2 - \omega^2}{1-v^2} } \frac{1}
{\sinh\hat{\beta}y}.
\label{shr}\end{equation}
From the expression (\ref{shr}) the scalar field $\phi$ is shown to
obey the following equation 
\begin{equation}
\pa^2\phi = \frac{1}{1-v^2}\frac{\sinh\phi}{\cosh^3\phi}
\left( \sinh^4\phi + 2\sinh^2\phi + \frac{1-v^2-\omega^2}{1-v^2}\right).
\label{phv}\end{equation}
When considering a special $\omega =0$ case, that is, 
the giant magnon on $AdS_2$, we have
\begin{equation}
\sinh\phi = \frac{1}{\sinh\frac{y}{\sqrt{1-v^2}} }
\end{equation}
and the eq. (\ref{phv}) implies that
the scalar field $\phi$ satisfies the sinh-Gordon equation
\begin{equation}
(\partial_{\tau}^2 - \partial_{\sigma}^2)2\phi = -\sinh 2\phi.
\end{equation}
Thus we see that there is a relation
between the giant magnon on $AdS_2$ and the sinh-Gordon soliton,
which corresponds to the map between the giant magnon on $S^2$
and the sine-Gordon soliton \cite{HM}.

Moreover, it is instructive to substitute the explicit solutions 
(\ref{sol}), (\ref{tat}) and (\ref{cor}) with $\varphi = \omega
(\tau - v\sigma)/(1-v^2)$ into the complex coordinates 
$Y_0$ and $Y_1$ in (\ref{coy}) and then obtain 
\begin{eqnarray}
Y_0 &=& \pm e^{i\tau}\left[ i \sqrt{1- \hat{\alpha}^2} -
\hat{\alpha}\coth\left(\hat{\beta}_0
\frac{y}{\sqrt{1-v^2}} \right)  \right], \nonumber \\
Y_1 &=& \mp\frac{\hat{\alpha}}{\sinh\left(
\hat{\beta}_0\frac{y}{\sqrt{1-v^2}}
 \right)} e^{i\sqrt{1- \hat{\beta}_0^2}(\tau -v\sigma)/\sqrt{1-v^2} },
\hspace{1cm}  \hat{\beta}_0 =\sqrt{\frac{1-v^2 - \omega^2}{1-v^2}},
\end{eqnarray}
whose signs correspond to $-\infty < y \le 0$ and $0 \le y < \infty$
respectively. This expression
for the string solution in $AdS_3$ looks similar to that expressed by
the two complex coordinates for the string solution in $R \times S^3$ 
corresponding to the dyonic giant magnon with two independent angular
momenta which was shown to be related with the charged soliton of
the complex sine-Gordon equation \cite{CDO}.

\section{Conclusion}

We have used the conformal gauge for the Polyakov action of strings in
$AdS_3 \times S^3$ to construct the three-spin giant magnon solution
with one spin in $AdS_3$ and two spins in $S^3$. 
By taking advantage of the explicit expression for the giant magnon
solution we have demonstrated a mapping between the giant magnon
on $AdS_2$ and the sinh-Gordon soliton. 

We have observed that the string configuration 
in $S^3$ makes an effect on the string motion in $AdS_3$ indirectly 
through the two Virasoro constraints.
From the Polyakov action of strings in $R \times S^3$ in the conformal 
gauge the static choice $t = \tau$ was used \cite{AFZ} to construct the
two-spin giant magnon solution in the SU(2) sector, while for the
three-spin giant magnon in $AdS_3 \times S^3$ this choice has not been
allowed such that the string time coordinate has a nontrivial dependence
on the worldsheet coordinates $\tau$ and $\sigma$. 
The arbitrary parameters $c_1$ and $c_2$ that characterize the time and
angle coordinates of string in $AdS_3$ have been chosen so as to 
satisfy the two Virasoro constraints. By regularizing the UV divergence
arising from the configuration of the string stretched to the
boundary of $AdS_3$, the dispersion relation of the three-spin giant
magnon has been obtained as the energy of a superposition of two
bound states of magnons. In the SU(2) subsector one bound state with
$J_2$ magnons has the total momentum which is given by 
the difference of the angle coordinate that is 
associated with infinite spin $J_1$, whereas
in the SL(2) subsector the other bound state with $|S_{\mathrm{reg}}|$
magnons has the total momentum which is specified by the difference
of the time coordinate that is associated with infinite energy $E$.


\begin{thebibliography}{99}
\bibitem{MGW} J.M. Maldacena, ``The large N limit of superconformal
field theories and supergravity,'' Adv. Theor. Math. Phys. \textbf{2}
(1998) 231 [hep-th/9711200]; S.S. Gubser, I.R. Klebanov and A.M. Polyakov,
``Gauge theory correlators from non-critical string theory,"
Phys. Lett. \textbf{B428} (1998) 105 [hep-th/9802109]; E. Witten, 
``Anti-de Sitter space and holography,"
Adv. Theor. Math. Phys. \textbf{2} (1998) 253 [hep-th/9802150].
\bibitem{BMN} D. Berenstein, J.M. Maldacena and H. Nastase, 
``Strings in flat space and pp waves from $\mathcal{N}$=4 super
Yang Mills," JHEP \textbf{04} (2002) 013 [hep-th/0202021].
\bibitem{GKP} S.S. Gubser, I.R. Klebanov and A.M. Polyakov,
``A semi-classical limit of the gauge/string correspondence,"
Nucl. Phys. \textbf{B636} (2002) 99 [hep-th/0204051].
\bibitem{FT} S. Frolov and A.A. Tseytlin, ``Semiclassical 
quantization of rotating superstring in $AdS_5\times S^5$," JHEP
\textbf{06} (2002) 007 [hep-th/0204226];
``Multi-spin string solutions in $AdS_5 \times S^5$," Nucl. Phys.
\textbf{B668} (2003) 77 [hep-th/0304255];
 ``Quantizing three-spin string solution in $AdS_5 \times S^5$,"
 JHEP \textbf{07} (2003) 016 [hep-th/0306130]; 
``Rotating string solutions: AdS/CFT duality in non-supersymmetric
sectors," Phys. Lett. \textbf{B570} (2003) 96 [hep-th/0306143].
\bibitem{AT} A.A. Tseytlin, ``Spinning strings and AdS/CFT duality,"
hep-th/0311139.
\bibitem{MZ} J.A. Minahan and K. Zarembo, ``The Bethe-ansatz for 
$\mathcal{N} =4$ super Yang-Mills," JHEP \textbf{03} (2003) 013 
[hep-th/0212208].
\bibitem{BKS} N. Beisert, C. Kristjansen and M. Staudacher, ``The 
dilatation operator of $\mathcal{N} =4$ super Yang-Mills theory," 
Nucl. Phys. \textbf{B664} (2003) 131 [hep-th/0303060];
N. Beisert and M. Staudacher, ``The $\mathcal{N}=4$ SYM integrable super
spin chain," Nucl. Phys. \textbf{B670} (2003) 439 [hep-th/0307042];
N. Beisert, V. Dippel and M. Staudacher, ``A novel long
range spin chain and planar $\mathcal{N} =4$ super Yang-Mills,"
JHEP \textbf{07} (2004) 075 [hep-th/0405001].
\bibitem{NB} N. Beisert, ``The dilatation operator of 
$\mathcal{N} = 4$ super Yang-Mills theory and integrability," 
Phys. Rept. \textbf{405} (2005) 1 [hep-th/0407277];
J. Plefka, ``Spinning strings and integrable spin chains 
in the AdS/CFT correspondence," hep-th/0507136.
\bibitem{BS} N. Beisert and M. Staudacher, ``Long-range PSU(2,2$|$4)
Bethe ansatze for gauge theory and strings," Nucl. Phys. \textbf{B727}
(2005) 1 [hep-th/0504190].
\bibitem{MK} M. Kruczenski, ``Spin chains and string theory,"
Phys. Rev. Lett. \textbf{93} (2004) 161602 [hep-th/0311203]; 
M. Kruczenski, A.V. Ryzhov and A.A. Tseytlin, ``Large spin limit
of $AdS_5 \times S^5$ string theory and low energy expansion
of ferromagnetic spin chains," Nucl. Phys. \textbf{B692} (2004) 3
[hep-th/0403120].
\bibitem{AAT} A.A. Tseytlin, ``Semiclassical strings and 
AdS/CFT,"  hep-th/0409296.
\bibitem{KMMZ} V.A. Kazakov, A. Marshakov, J.A. Minahan and K. Zarembo,
``Classical/quantum integrability in AdS/CFT," JHEP \textbf{05}
(2004) 024 [hep-th/0402207];
N. Beisert, V.A. Kazakov and K. Sakai, ``Algebraic curve for the SO(6)
sector of AdS/CFT," Commun. Math. Phys. \textbf{263} (2006) 611 
[hep-th/0410253];
N. Dorey and B. Vicedo, ``On the dynamics of finite-gap solutions in
classical string theory," JHEP \textbf{07} (2006) 014 [hep-th/0601194].
\bibitem{KZ} K. Zarembo, ``Semiclassical Bethe ansatz and AdS/CFT," 
Comptes Rendus Physique \textbf{5} (2004) 1081 [hep-th/0411191].
\bibitem{HM} D.M. Hofman and J.M. Maldacena, ``Giant magnons,"
hep-th/0604135.
\bibitem{SR} S. Ryang, ``Wound and rotating strings in $AdS_5 \times
S^5$," JHEP \textbf{08} (2005) 047 [hep-th/0503239].
\bibitem{MKS} M. Kruczenski, ``Spiky strings and single trace
operators in gauge theories," JHEP \textbf{08} (2005) 014
[hep-th/0410226].
\bibitem{NBT} N. Beisert, ``The su(2$|$2) dynamic S-matrix,"
hep-th/0511082.
\bibitem{KP} K. Pohlmeyer, ``Integrable Hamiltonian systems and 
interactions through quadratic constraints," Commun. Math. Phys.
\textbf{46} (1976) 207.
\bibitem{AM} A. Mikhailov, ``An action variable of the sine-Gordon model,"
hep-th/0504035;
``A nonlocal Poisson bracket of the sine-Gordon model," 
hep--th/0511069.
\bibitem{AFS} G. Arutyunov, S. Frolov and M. Staudacher, ``Bethe
ansatz for quantum strings," JHEP \textbf{10} (2004)
016 [hep-th/0406256].
\bibitem{ND} N. Dorey, ``Magnon bound states and the AdS/CFT 
correspondence," hep-th/0604175.
\bibitem{CDO} H.Y. Chen, N. Dorey and K. Okamura, ``Dyonic giant
magnons," hep-th/0605155.
\bibitem{AFZ} G. Arutyunov, S. Frolov and M. Zamaklar, ``Finite-size 
effects from giant magnons," hep-th/0606126.
\bibitem{MTT} J.A. Minahan, A. Tirziu and A.A. Tseytlin, ``Infinite
spin limit of semiclassical string states," JHEP \textbf{08} (2006)
049 [hep-th/0606145].
\bibitem{CGK} C.S. Chu, G. Georgiou and V.V. Khoze, ``Magnons,
classical strings and beta-deformations," hep-th/0606220.
\bibitem{BR} N.P. Bobev and R.C. Rashkov, ``Multispin giant magnons,"
Phsy. Rev. \textbf{D74} (2006) 046011 [hep-th/0607018].
\bibitem{OS} K. Okamura and R. Suzuki, ``A perspective on classical 
strings from complex sine-Gordon solitons," hep-th/0609026.
\bibitem{ZM} V.E. Zakharov and A.V. Mikhailov, ``Relativistically
invariant two-dimensional models in field theory integrable by the
inverse problem technique (in Russian)," Sov. Phys. JETP \textbf{47}
(1978) 1017 [Zh. Eksp. Teor. Fiz. \textbf{74} (1978) 1953 ];
``On integrability of classical spinor models in two-dimensional
space-time," Commun. Math. Phys. \textbf{74} (1980) 21;
J.P. Harnad, Y. Saint Aubin and S. Shnider, ``Backlund transformations
for nonlinear sigma models with values in Riemannian symmetric spaces,"
Commun. Math. Phys. \textbf{92} (1984) 329.
\bibitem{SV} M. Spradlin and A. Volovich, ``Dressing the giant
magnon," hep-th/0607009.
\bibitem{KRT} M. Kruczenski, J. Russo and A.A. Tseytlin, ``Spiky strings
and giant magnons on $S^5$," hep-th/0607044.
\bibitem{ART} G. Arutyunov, J. Russo and A.A. Tseytlin, ``Spinning 
strings in $AdS_5 \times S^5$: new integrable system relations," 
Phys. Rev. \textbf{D69} (2004) 086009 [hep-th/0311004].
\bibitem{CR} H.Y. Chen, N. Dorey and K. Okamura, ``On the scattering
of magnon boundstates," hep-th/0608047;
R. Roiban, ``Magnon bound-state scattering in gauge and string theory,"
hep-th/0608049.
\bibitem{WH} W.H. Huang, ``Giant magnons under NS-NS and Melvin fields,"
hep-th/0607161.

\end{thebibliography}
\end{document}